\documentstyle[aps,prl,graphicx,twocolumn]{revtex}

\newcommand{\ket}[1]{|#1\rangle}

\newcommand{\bra}[1]{\langle#1|}

\newcommand{\Exp}[1]{{\mathrm{e}^{#1}}}

\newcommand{\bm}[1]{{\mathbf{#1}}}

\newcommand{\tm}[1]{{\mathrm{#1}}}

\newcommand{\tim}[1]{{\mbox{#1}}}

\newcommand{\lkl}{\left(}

\newcommand{\rkl}{\right)}

\newcommand{\re}{\mathrm{Re}}

\newcommand{\im}{\mathrm{Im}}

\newcommand{\I}{\mathrm{i}}

\newcommand{\hc}{\mbox{h.c.}}

\begin{document}

\draft

\title{Monitoring the Dipole-Dipole Interaction via Quantum Jumps of Individual Atoms}

\author{C. Skornia$^{\ast\dagger}$, J.von Zanthier$^\ast$, 
  G.S. Agarwal$^{\ast\ddagger}$, E. Werner$^{\dagger}$, H. Walther$^\ast$}

\address{$^\ast$ Max-Planck-Institut f{\"u}r Quantenoptik and Sektion
  Physik der LMU M{\"u}nchen, 
  D-85748 Garching, Germany\protect\\  
$^\dagger$ Institut f{\"u}r Theoretische Physik, Universit{\"a}t Regensburg, 
  D-93040 Regensburg, Germany\protect\\  $^\ddagger$ Physical Research Laboratory, Navrangpura, Ahmedabad-380 009, India}

\date{\today}

\maketitle

\begin{abstract}

The emission characteristics in the fluorescence of two laser-driven dipole-dipole-interacting three level atoms is investigated. When the light from both atoms is detected separately a correlation of the emission processes is observed in dependence of the dipole-dipole interaction. This opens the possibility to investigate the dipole-dipole interaction through the emission behavior. We present Monte-Carlo simulations which are in good agreement with the analytic solutions.

\end{abstract}






The problem of the  influence of a neighboring  atom on atomic emission behavior \cite{cook85,cohen-tannoudji86,cook90,nagourney86} has  been investigated  since a very long time. Especially the problem of the  correlation of the emission of two neighboring atoms has found quite some interest \cite{toschek86,bergquist86,hulet88}. This  is a complex issue and the answer is very much dependent on various system parameters such as the strength of the pumping field, the life times and the wavelengths of the different transitions involved.  An early theoretical calculation concluded that the dipole-dipole interaction between adjacent atoms is irrelevant for quantum jumps within a 3-level-system under the  usual experimental conditions where the  Rabi frequency of the pump field is large compared to all other rates in the problem \cite{lewenstein88,agarwal77}. Those results have been confirmed by quantum Monte-Carlo calculations \cite{carmichael99,dalibard93,hegerfeldt93}, \cite{skornia00}. In  this  paper we reconsider the problem of dipole-dipole interaction  and investigate especially the correlations in the emissions of two neighboring atoms which are observed individually.\\

\begin{figure}[htbp] 
\begin{center}    
\includegraphics[height=4.8cm]{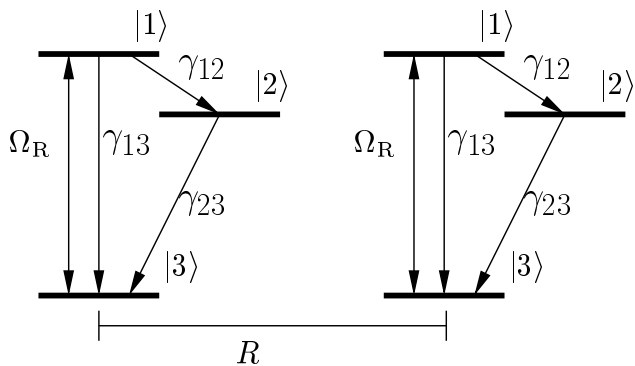}    
\caption{Two dipole-dipole-interacting 3-level-atoms, separated by  a distance $R$ in a laser field resonant to the transition      $\ket{1} \leftrightarrow \ket{3}$}    
\label{fig:setup}
  
\end{center}
\end{figure}

The arrangement is shown schematically in Fig.~\ref{fig:setup}. We consider 2 identical  nearby atoms in a trap, each with levels $\ket{1}$, $\ket{2}$ and
$\ket{3}$. One of the atoms (say atom 1) is initially prepared, e.g.  by pulsed excitation \cite{nagerl99}, in a metastable state $\ket{2}$.  A resonant cw pump interacts with both atoms. The initial state  of atom 1 is such that it can start interacting with the pump either due to decay to the state $\ket{3}$  by emission of a photon or via excitation to the state $\ket{1}$ as a result of the dipole-dipole interaction. In the second  case atom 2 goes to the state $\ket{2}$. In our analysis we assume that the level spacing $\ket{1} \leftrightarrow \ket{2}$ is close enough,  so that the dipole-dipole interaction is effective only on this transition. The transitions $\ket{1} \leftrightarrow \ket{3}$  and  $\ket{2} \leftrightarrow  \ket{3}$ are supposed to lie in the optical domain, where the distance between two atoms is assumed to be much larger than the wavelengths of the  corresponding transitions.  For example in In$^+$ the wavelength on the $\ket{1} \leftrightarrow \ket{2}$ corresponds to $9.3 \mu$m which could be two to three times the distance between two trapped ions \cite{peik99}. We further assume that the  fluorescence from each atom can be resolved individually by two distinct detectors 1 and 2.  In what follow let us work in the limit

\begin{equation} 
\label{eq:assumption}  
\Omega_{R} > \gamma^{13} \gg \gamma^{12}, \gamma^{23}
\end{equation}

In our prepared system, detector 1 will not detect any fluorescence while atom 2 is   fluorescing on the $\ket{1} \leftrightarrow  \ket{3}$ transition until the dipole-dipole interaction on the transition $\ket{1}  \leftrightarrow \ket{2}$ brings  atom 1 towards the cycling  transition $\ket{1} \leftrightarrow \ket{3}$ and atom 2 towards the  metastable state $\ket{2}$.  In  this case detector 1 is switched  on and at the same time detector 2 is switched off.  Detector 2 will remain off until either the atom 2  makes a transition from the state $\ket{2}$ to $\ket{3}$ by spontaneous emission or again to the state $\ket{1}$ by the dipole-dipole interaction. If  atom 2 goes to the state $\ket{3}$, both the detectors will be on. If on the other hand atom 2 goes to the state $\ket{1}$ by dipole-dipole interaction,  then again detector 1 will  switch off. Under the above inequality the dipole-dipole interaction is more  probable and hence the chance for both detectors 1 and  2 to switch  on at the same time is rare. We thus conclude that the fluorescence records of the detectors 1  and 2 will be complementary. This complementary record of fluorescence is a clear signature of the dipole-dipole interaction between the  two atoms.  The key ingredients in the above argument are (a) the inequality  (1), (b) preparation  of one  of the atoms in the metastable state $\ket{2}$, (c) capability to resolve the fluorescence from individual atoms which are separated by a distance much bigger than the wavelength of the strong transition. In what follows we will present results from the quantum Monte-Carlo simulation of the fluorescence from the two 3-level-atoms. These simulations validate the above ideas.\\


The two identical atoms are assumed to be at fixed positions $\bm{r}_i$ , and the field outside the laser-beam in the vacuum state.  The dipole matrix elements of the atoms are defined as  $\bm{d}_{ij} :=   e\bra{i}x\ket{j}$ and the  atomic operators for the $i$-th     atoms as $\sigma_i^{\tm{nm}} := \ket{\tm{n}}_i\bra{\tm{m}}$. The interaction hamiltonian for two dipole-dipole
interacting three-level-systems reads \cite{agarwal74}

\begin{equation}
\label{eq:Heasy}  
H = \hbar  \sum_{\tm{n},\tm{m}}\sum_{\bm{k}\bm{s}}\lkl g_{\bm{k}\bm{s}}a_{\bm{k}\bm{s}}  \sigma^{\tm{nm}}_{\bm{k}} + \hc \rkl
\end{equation}

with

\begin{eqnarray}
\label{eq:Atomic-Operators-2}  
\sigma^{\tm{nm}}_{\bm{k}}&:=& \sum_i \sigma^{\tm{nm}}_i \Exp{\I \bm{k}\cdot\bm{r}_i} \\  \tim{and} \quad   g_{\bm{k}\bm{s}} &:=& - \I \sqrt{\frac{\omega}{2 \hbar \epsilon_0 V}}   \lkl \bm{d} \cdot \bm{\epsilon}_{\bm{k}\bm{s}} \rkl.
\end{eqnarray}

For the time steps  $\Delta t$ used in the simulations we need the time-development of the system under the condition that no photon is emitted in the time between  $t=0$ and $t=\Delta t$. This is described by a non-hermitian `conditional hamiltonian', including the atom laser interaction, which is found to be \cite{carmichael99,dalibard93,hegerfeldt93,beige99}

\begin{eqnarray} 
\label{eq:Hcond}  
H_{\tim{\footnotesize cond}} &=& \frac{\hbar}{\I}\left[  \sum_{{\tm{n,m} = 1 \atop \tm{ n}<\tm{m}}}^3 \sum_{\tm{i},\tm{j}=1}^2\gamma_{\tm{ij}}^{\tm{nm}}  \sigma^{\tm{nm}}_{\tm{i}}\sigma^{\tm{mn}}_{\tm{j}} \right. \nonumber\\ 
&&\hspace{1.25cm}\left. + \sum_{\tm{i}=1}^2\I\Omega_R\lkl \sigma_{\tm{i}}^{13}  +\sigma_{\tm{i}}^{31}\rkl \right].
\end{eqnarray}

Here  $2 \gamma_{\tm{ij}}^{\tm{nm}}$  is the  Einstein  $A$-coefficient for the
transition $\ket{n} \leftrightarrow \ket{m}$ for  $i=j$ and $2 \Omega_R$ is  the Rabi  frequency.   For $i  \neq j$,  $\gamma_{\tm{ij}}^{\tm{nm}}$  is the complex parameter  which describes the  strength of the dipole-dipole interaction, where
we  have      made the  rotating-wave-approximation     in   the  derivation  of (\ref{eq:Hcond}).  In    the Markov   approximation  they   can   be  calculated
analytically \cite{agarwal74}:

\begin{eqnarray}  
\label{eq:2}  
\gamma_{\tm{12}}^{\tm{nm}} &=& \frac{3}{2}\gamma_{\tm{11}}^{\tm{nm}}\Exp{\I k_{\tm{nm}}r}  \left[ \frac{1}{\I k_{\tm{nm}}r}\lkl1-\cos^2\theta_{\tm{nm}}\rkl  \right. \nonumber \\ &&\left. + \lkl     \frac{1}{(\I k_{\tm{nm}}r)^2} +  \frac{1}{(\I k_{\tm{nm}}r)^3}    \rkl \lkl 1-3\cos^2\theta_{\tm{nm}} \rkl  \right]
\end{eqnarray}

Here  $r =  |\bm{r}_1-\bm{r}_2|$ is the  distance  between  the two  atoms and $\theta_{\tm{nm}}$ the angle between $\bm{d}_{\tm{nm}}$ and $\bm{r}_1-\bm{r}_2$. For  the further calculations we define $\gamma^{\tm{nm}}=\gamma_{\tm{11}}^{\tm{nm}} = \gamma_{\tm{22}}^{\tm{nm}}$, $\gamma^{\tm{nm}}_{\tm{dd}}  :=  \re\lkl \gamma_{\tm{12}}^{\tm{nm}}\rkl$  and $\Omega^{\tm{nm}}_{\tm{dd}} := \im\lkl \gamma_{\tm{12}}^{\tm{nm}}\rkl$

\begin{figure}[htbp]  
\begin{center}    
\includegraphics[height=4.5cm]{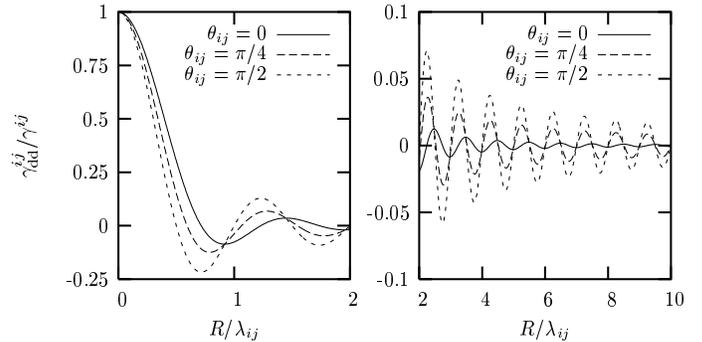}    
\caption{Real part of dipole-dipole interaction parameter $\gamma^{\tm{nm}}_{\tm{dd}}$ in units of $\gamma^{\tm{nm}}$}    
\label{fig:gamr}  
\end{center}
\end{figure}

\begin{figure}[htbp]  
\begin{center}    
\includegraphics[height=4.5cm]{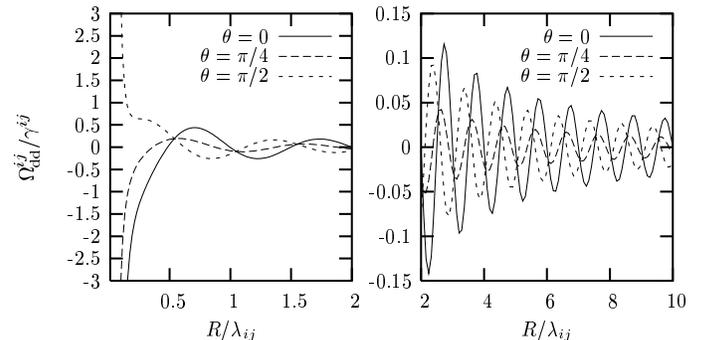}    
\caption{Imaginary part of dipole-dipole interaction parameter $\Omega^{\tm{nm}}_{\tm{dd}}$ in units of $\gamma^{\tm{nm}}$}    
\label{fig:Omr}  
\end{center}
\end{figure}

Their spatial dependence is shown in  figure \ref{fig:gamr} and \ref{fig:Omr}. Obviously  $\gamma^{\tm{nm}}_{\tm{dd}}$ and $\Omega^{\tm{nm}}_{\tm{dd}}$ tend to zero  for $r\to\infty$, but for  $r\to 0$ $\Omega^{\tm{nm}}_{\tm{dd}}$ diverges. This means that $\Omega^{\tm{nm}}(r)$ can have important consequences for the evolution of the system for $r < \lambda_{nm}$. By use of the
Monte-Carlo simulations we indeed show, that this is the case for the quantum-jump-behavior of the system.

The evolution of the state vector between two consecutive emissions is given by \cite{carmichael99,dalibard93,hegerfeldt93,beige99}

\begin{equation}  
\label{eq:evol}  
\ket{\Psi(t+\Delta t)}= \lkl 1 - \frac{\I}{\hbar} H_{\tim{\footnotesize cond}}\Delta t\rkl \ket{\Psi(t)}
\end{equation}

where we assume that  $\Delta t$ is  much larger than the inverse optical frequencies but much smaller than all the atomic decay times.\\


To show  that the dipole-dipole interaction can have important consequences for the quantum-jump-behavior of the system we detect the  fluorescence light of the two atoms separately with the detectors 1  and 2, which are modeled by the operators

\begin{equation}  
\label{eq:detectors}  
D_{\tm{i}}=\ket{3}_{\tm{i}}\bra{1}
\end{equation}

where  the  probability $P_{\tm{i}}$ for  a measurement with the detector i is proportional to the excitation probability of the i-th atom. We find:

\begin{equation}  
\label{eq:Probs}  
P_{\tm{i}}=2\gamma^{13}\langle D_1 \rangle
\end{equation}

After each measurement we have to continue with the reduced state $\ket{\Psi}_{\tim{\footnotesize red}}= \frac1N D_{\tm{i}}\ket{\Psi}$, with N being a normalization constant.\\

Next we focus on the dipole-dipole interaction on the transition $\ket{1} \leftrightarrow  \ket{2}$, where the $r$-dependent factor $\frac{\lambda_{12}}{r}$ is largest. In the time evolution  of the system which is given by equation (\ref{eq:evol}) the  interaction parameter $\gamma_{12}^{12}$ appears with the operators

\begin{equation}  
\label{eq:entangle}  
\sigma_1^{12}\sigma_2^{21}+\sigma_2^{12}\sigma_1^{21}.
\end{equation}

If the probability amplitude for the system being in state $\ket{1,2}=\ket{1}_1\otimes\ket{2}_2$ is not zero for the initial state the probability amplitude of being in  state $\ket{2,1}$ is not zero one time step further. Consequently the  dipole-dipole-interaction drives the system from a non-entangled state to an entangled state  \cite{jaksch00}. This is a non-classical effect, which we can detect with our setup. The  relevant time scale of this process is $\frac{1}{\gamma_{12}^{12}}$  as is clear from equation (\ref{eq:Hcond}). To see the effect, on atom has to be in state $\ket{2}$, which can be prepared by an appropriate laser pulse, while the other one must have a non zero probability of being in state $\ket{1}$ what is achieved by the driving laser. The initial state for the simulation is therefore $\ket{1,2}$. It is evident, that every `click' of one of our detectors destroys the entanglement and one starts again with a separable state. There are no problems  with decoherence effects, as no coherence needs to be reserved longer than  the lifetime of the upper  state $\ket{1}$.\\



If we  compare the time scales and probabilities in the regime (\ref{eq:assumption}) it is clear that atom 1 shows fluorescence, i.e. detector 1 clicks, while atom 2 is dark. There are two possible decays which change the fluorescence behavior of the system (a) atom 1 falls  to $\ket{2}$ and both detectors do not detect any further photons (b) atom 2 falls  to $\ket{3}$ and both atoms show fluorescence at the same time. There  is also a probability for a third event which is the most interesting in this context. As mentioned above with every time step  where we have a non zero probability amplitude of the system being in state $\ket{1,2}$ we get a non zero probability  amplitude for being in state $\ket{2,1}$, due to the dipole-dipole interaction. If this transition occurs, atom 2 gets excited to  state $\ket{1}$ from where it starts to cycle on the $\ket{1}  \leftrightarrow \ket{3}$ transition and to emit fluorescence photons, while atom 1 ends up in the metastable state $\ket{2}$ and gets dark.  Finally we have three possible changes of the detection rates, which are clearly distinguishable.  As we are interested only in the third, we do not count the times where both atoms are bright or dark. We assume that in this case we prepare our initial state again. We are then in a  situation where always one atom is in  state $\ket{2}$.  Consequently the  dipole-dipole interaction on the transition  $\ket{1}\leftrightarrow \ket{3}$ has  no effect as the corresponding operators in the time evolution are zero.

\begin{figure}[htbp]  
\begin{center}    
\includegraphics[height=5cm]{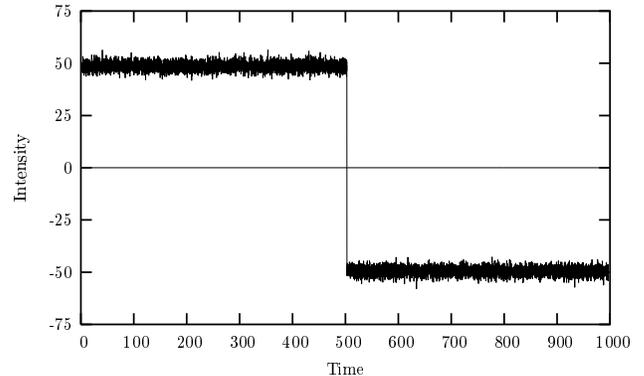}    
\label{fig:flip}    
\caption{Flip in the detection rates of the two detectors. The time interval for the summation of detections is $\frac{50}{\gamma^{13}}$ for the Rabi-frequency $\Omega_R = 8$. The emissions of atom 1 are plotted with positive, those of atom 2 with negative sign, so the plot shows an event where atom 1 gets dark and atom 2 bright at the same time .}    
\label{fig:flip}  
\end{center}
\end{figure}

\begin{figure}[htbp]  
\begin{center}    
\hspace{-18pt}
\includegraphics[height=4.6cm]{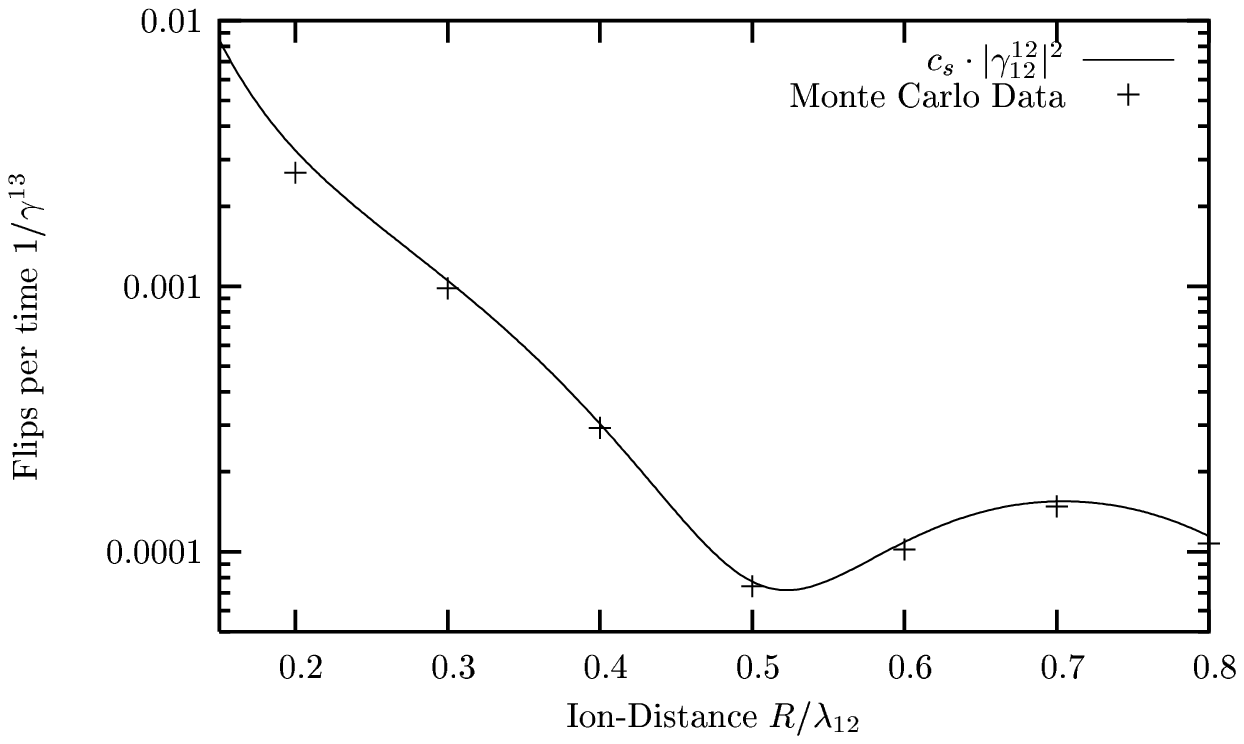}\\[2ex]
\hspace{-18pt}
\includegraphics[height=4.6cm]{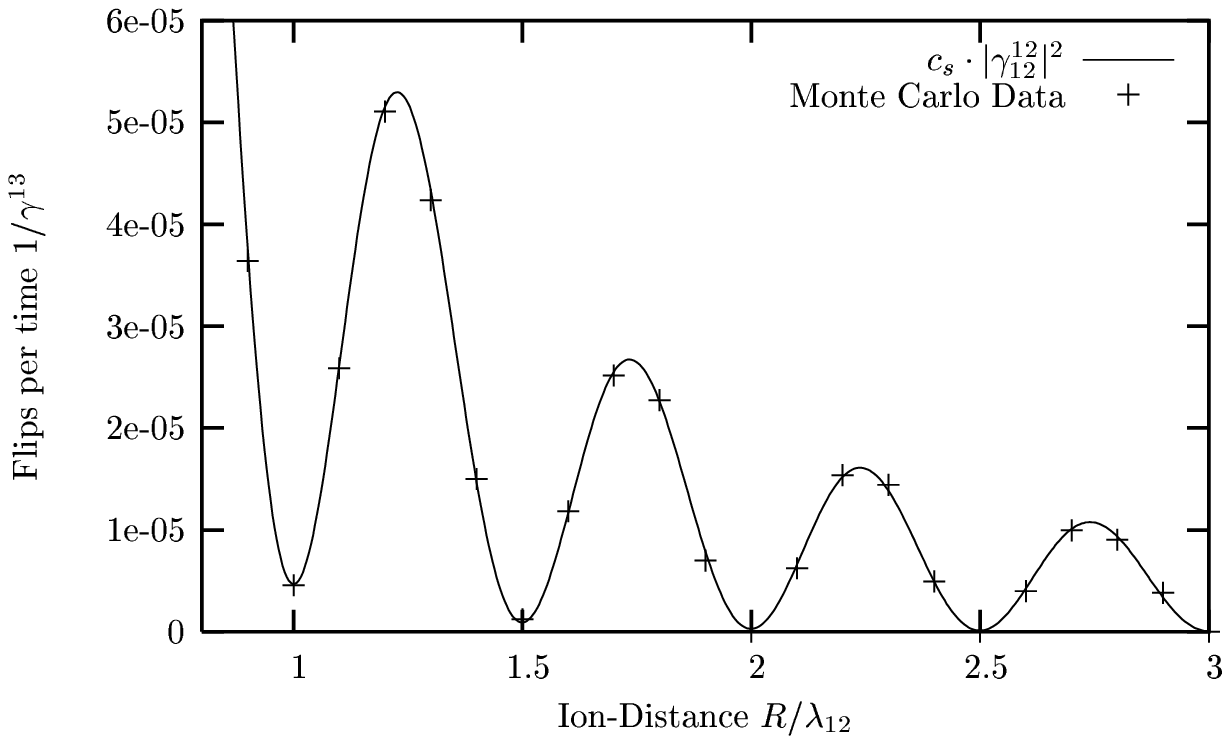}    
\caption{Monte-Carlo data of Flips per time compared to $|\gamma_{\tm{12}}^{\tm{12}}|^2$ for $\frac{\gamma^{\tm{12}}} {\gamma^{\tm{13}}}=2 \cdot 10^{-2}$ with $c_s =2$ and $\Omega_R = 8$.}   
\label{fig:data1}  
\end{center}
\end{figure}

\begin{figure}[htbp]  
\begin{center}    
\hspace{-18pt}
\includegraphics[height=4.6cm]{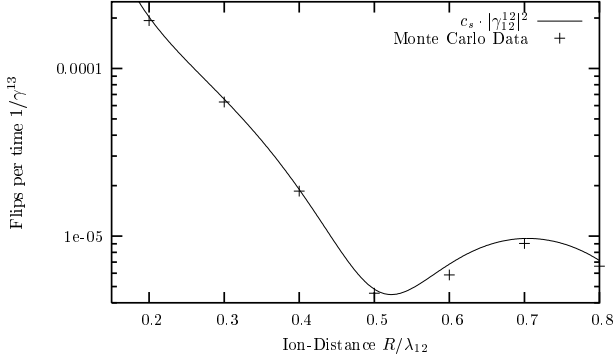}\\[2ex]   
\hspace{-18pt}
\includegraphics[height=4.6cm]{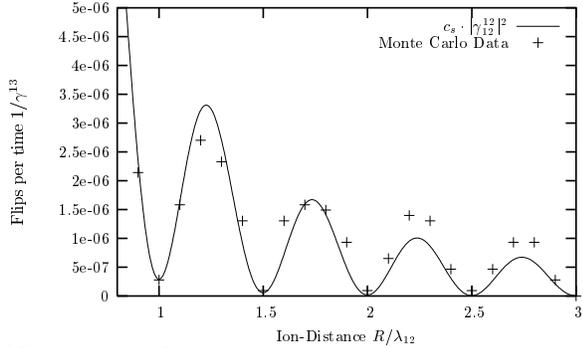}    
\caption{Monte-Carlo data of Flips per time compared to $|\gamma_{\tm{12}}^{\tm{12}}|^2$ for $\frac{\gamma^{\tm{12}}}{\gamma^{\tm{13}}}=5 \cdot 10^{-3}$ with $c_s =2$ and $\Omega_R = 8$} \label{fig:data2}  
\end{center}
\end{figure}

The flip in the  detection of fluorescence (Fig. \ref{fig:flip}) is a  clear signature of the dipole-dipole interaction of  the  system  and shows how the quantum jump statistic can be modified  by cooperativity.  Note that the effect of dipole-dipole interaction is not observable if  we do not distinguish between the fluorescence of the two atoms \cite{skornia00}. To  see how  this behavior is connected  to the interaction parameter $\gamma_{\tm{12}}^{\tm{12}}$ we count the flips  per unit time for different atom distances of interest (from $r=0.1\lambda_{12}$  to $r=3\lambda_{12}$) and find, up to a scaling factor, very good  functional agreement with $|\gamma_{\tm{12}}^{\tm{12}}|^2$ (Fig.  \ref{fig:data1}, \ref{fig:data2}).

If we further analyze our  data for different ratios of  the decay rates $\gamma^{\tm{12}}$ and $\gamma^{\tm{13}}$ we find that the remaining scaling factor $c_s$  is  nearly constant over reasonable magnitudes.

\begin{figure}[htbp]  
\begin{center}    
\fbox{\rule{0cm}{5cm}\rule{7cm}{0cm}}    
\vspace{0.5cm}    
\hspace{-3cm}
\includegraphics[height=4.8cm]{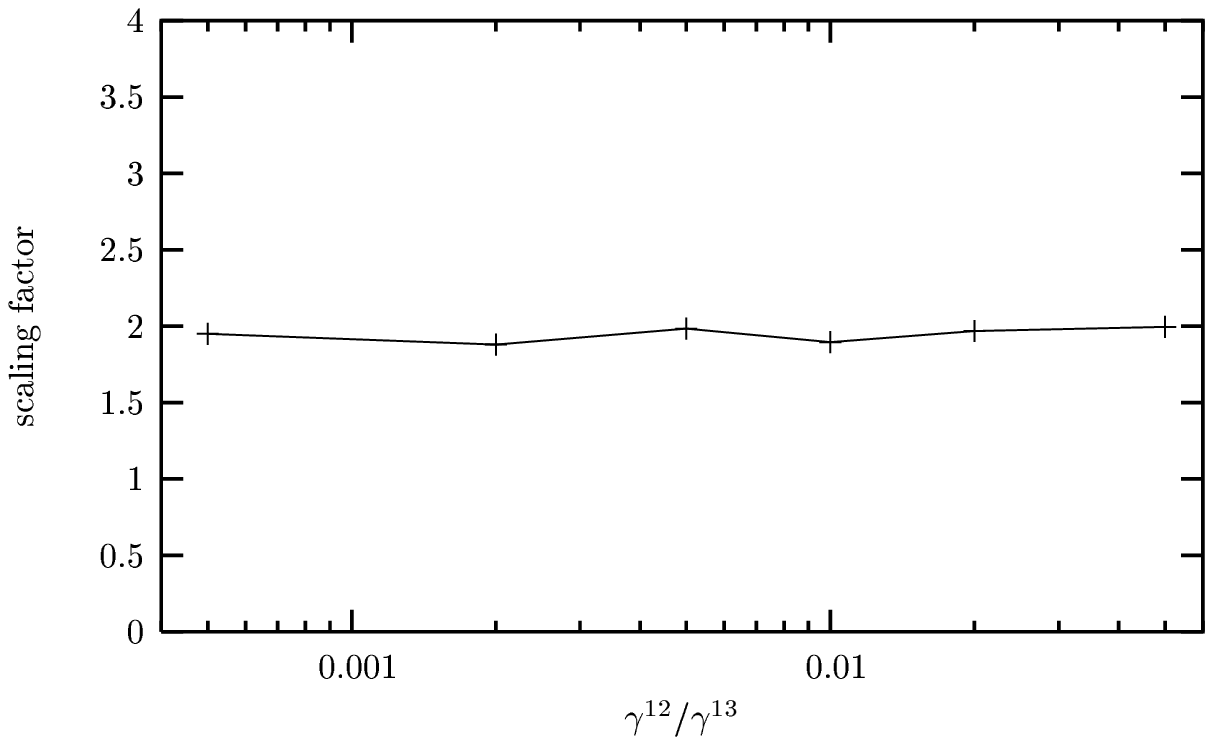}    
\caption{Scaling factor for different ratios of $\gamma^{\tm{12}}$ to $\gamma^{\tm{13}}$  with $\Omega_R = 8$. We  find  this factor by a least-square fit of the Monte-Carlo Data to the values of $|\gamma_{\tm{12}}^{\tm{12}}|^2$}    
\label{fig:scalfac}
\end{center}
\end{figure}

The result is shown in  Fig.~\ref{fig:scalfac}. This shows that indeed the flip rate is a consequence of the dipole-dipole interaction and that other parameters than $\gamma_{12}^{12}$ do not intervene.


In conclusion, we have outlined one solution to the problem of detecting the dipole-dipole interaction in the phenomena of quantum jumps. For that purpose we analyzed a system of two dipole-dipole interacting 3-level-systems and proposed, in contrast to earlier calculations \cite{lewenstein88,agarwal77,beige99}, a detection scheme, where we distinguish between the two  atoms in order to monitor the consequences of this interaction. Within this scheme there are no problems related to decoherence, as no coherences need to be preserved  over timescales longer than the shortest decay time of the system. Furthermore the  dependence of the flip rate on the spatial variation of the dipole-dipole interaction has been analyzed.

\end{document}